\documentclass[aps,pra, twocolumn,a4paper,showpacs]{revtex4}
\usepackage{amsmath}
\usepackage{bm}
\usepackage{graphicx}

\def\BE {\begin{equation}}
\def\EE {\end{equation}}
\def\BEA {\begin{eqnarray}}
\def\EEA {\end{eqnarray}}
\def\BES {\begin{subequations}}
\def\EES {\end{subequations}}
\def\BA {\begin{array}}
\def\EA {\end{array}}
\def\NN {\nonumber}
\def\NN {\nonumber}
\def\ep {\varepsilon}

\begin{document}

\title{Analog Grover search by adiabatic passage in a cavity-laser-atom system}

\author{D. Daems}
\affiliation{QuIC, Ecole Polytechnique,
Universit\'e Libre de Bruxelles, 1050 Bruxelles, Belgium}
\email{ddaems@ulb.ac.be}

\author{S. Gu\'erin}
\affiliation{Institut Carnot de Bourgogne UMR 5209 CNRS,
Universit\'e de Bourgogne, BP 47870, 21078 Dijon, France}
\email{sguerin@u-bourgogne.fr}

\abstract{A physical implementation of the adiabatic Grover search
is theoretically investigated in a system of $N$ identical
three-level atoms trapped in a single mode cavity. Some of the
atoms are marked through the presence of an energy gap between
their two ground states.  The search is controlled by two
partially delayed lasers which allow a deterministic adiabatic
transfer from an initially entangled state to the marked states.
Pulse schemes are proposed to satisfy the Grover speedup either
exactly or approximately, and the success rate of the search is
calculated.} }

 \maketitle


\section{Introduction}   
Search problems can be expressed as finding a set of marked items in an unsorted list.
The Grover algorithm \cite{grover} achieves this task quadratically faster than a classical algorithm which examines items one by one.
It has become a paradigm of quantum computation based on quantum circuits.
The case of two qubits, corresponding to a four-element search, has been demonstrated experimentally in various settings.
NMR experiments \cite{nmr1,nmr2,nmr3} used the spin states of $^1$H and/or $^{13}$C in a magnetic field as qubits while radio-frequency fields and spin-spin couplings between the nuclei were used to implement the quantum logic gates.
In  linear optics techniques, the individual qubits are represented by  different polarization or spatial-mode degrees of freedom while the computation is achieved through essentially a complicated interferometer \cite{optics1} or a one-way quantum computer   \cite{optics2}.
Trapped ions systems  \cite{ions}, which present the advantage of being scalable, rely on a number of optical and microwaves sources to control, entangle and measure the qubits which are represented by the ground state hyperfine levels of two trapped atomic ions.
There have also been proposals of experimental implementations using cavity QED where the quantum
gate dynamics is provided by a cavity-assisted collision
\cite{qed1} or by a strong resonant classical field \cite{qed2}.

A different approach to quantum computation  consists in the controlled evolution of a system obeying the Schr\"odinger equation with a Hamiltonian designed to solve a specified problem.
It was pioneered by Farhi and Gutmann \cite{farhi}
who considered a time continuous version of the Grover algorithm.
A given Hamiltonian features a marked state whose energy differs from that of the $N-1$ unmarked ones.
A driving Hamiltonian is then constructed, without any knowledge of the solution, to produce a Rabi-like half-cycle which populates the marked state in a time scaling as $N^{1/2}$, thereby exhibiting the quadratic speedup.
 An experimental realization of this analog Grover
algorithm has been performed by NMR \cite{nmr4} in a setting where
a quadrupolar coupling  makes a spin 3/2 nucleus a two-qubit
system ($N$=4).

Adiabatic processes offer many advantages, in particular, the high degree of population transfer and some robustness with respect to fluctuations of the control fields or imperfect knowledge of the model.
Adiabatic versions of the time continuous Grover algorithm have
been constructed in abstract form \cite{adiabatic1,adiabatic2,roland}.
An ad hoc Hamiltonian
connects adiabatically the initial ground superposition to the
marked state through an effective two-state avoided crossing.
 It has been shown \cite{roland} that the transfer to the marked
state in a time growing as $N^{1/2}$ requires a specific
time-dependent sweeping of the parameter controlling the Hamiltonian.
We have recently proposed a physical implementation of such an
adiabatic search using three-state atoms  trapped  in a QED cavity
and driven by laser fields \cite{PRL}. This scheme leads to an effective
three-state dynamics which is closely related to  stimulated Raman
adiabatic passage (STIRAP) \cite{STIRAP}. We have determined
analytically the shapes of the adiabatic pulses that are required to lead to
a quadratic speedup of the search.

The present work contains a detailed description of the techniques and results announced in Ref. \cite{PRL}.
We consider the more general case of search problems which have more than one solution.
The dynamics of the collective marked, unmarked and excited states introduced below is solved exactly under local adiabatic conditions for which the scaling is determined analytically.
Furthermore, we study numerically the
robustness of the Grover search in this system using pulses
which are easily generated in practice (typically Gaussian pulses with plateaus).
We identify the pertaining conditions which give rise to a
speedup of the search which is only approximately quadratic in this case.

The paper is organized as follows.
In Sec. I, we describe the cavity-laser-atom system, introduce the relevant collective states and derive the effective Hamiltonian.
Section III is devoted to the analysis of the adiabatic conditions leading to a Grover search with an exact or approximate quadratic speedup.
The two types of pulses schemes are illustrated and discussed in Sec. IV while
the conclusions are given in Sec. V.

\begin{center}
\begin{figure}
\includegraphics[scale=0.55]{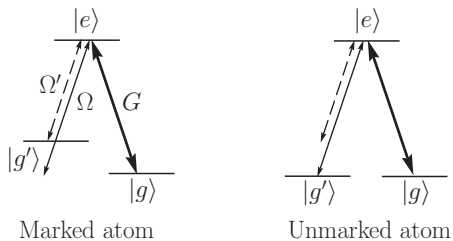}
\caption{Linkage pattern for the individual atoms. The unmarked
atoms have two degenerate ground states $|g\rangle$ and
$|g'\rangle$. The marked atoms feature the state $|g'\rangle$
shifted. The laser of Rabi frequency $\Omega'$ (resp. $\Omega$) is
resonant with the $g'-e$ transition for the marked (resp.
unmarked) atoms. The cavity of Rabi frequency $G$ is resonant with
the $g-e$ transition.} \label{atoms}
\end{figure}
\end{center}

\section{Cavity-laser-atom system}

\subsection{Description}

We use an ensemble of $N$ identical three-level atoms trapped in a
single-mode cavity of coupling frequency $G$, and driven by two
lasers of Rabi frequencies $\Omega$ and $\Omega'$. The key of the
search process is that all the atoms have to be entangled
initially as defined below. The atoms have a $\Lambda$
configuration with two ground Zeeman states $|g\rangle$ and
$|g'\rangle$, coupled to the excited state $|e\rangle$ by
respectively the cavity and the two lasers (see Fig. 1). The
states $|g'\rangle$ of all the atoms are considered as the
database. The state $|g\rangle$ stands here to couple all the
atoms together by exchanging a single photon in the cavity. It
allows us to consider the unmarked collective state, defined as
the normalized sum of all the possible states $|g'_j\rangle$ for
all the unmarked atoms in state $|g\rangle$ except the
$j^{\hbox{\scriptsize th}}$ in $|g'\rangle$. The state
$|g'\rangle$ of any marked atom is shifted, for instance by a
magnetic field, such that the transition $g'-e$ is resonant with
the laser $\Omega'$ (resp. $\Omega$) for a marked atom (resp.
unmarked atom). In this scheme, the magnetic field should be
atomic-selective and the atoms should thus be fixed and form a
register (of one or two dimension) in the cavity. We start from
the initial entangled state $
|g^\prime,0\rangle\equiv(1/\sqrt{N})\sum_{j=1}^N|g'_j,0\rangle $
featuring a collective superposition of both types of ground states. Such a
state can be prepared for instance before the marking of the atom
using the stimulated Raman adiabatic passage (STIRAP) technique,
exactly as shown by Fleischhauer and coworkers to store
single-photon quantum states \cite{Fleisch}. In their scheme, a
single-photon wave packet enters the cavity through a mirror while
the resonant laser $\Omega$ is on and strong as $\Omega\gg
G\sqrt{N}$, and the atoms are all in their ground state
$g\rangle$. This generates the state denoted $|g,1\rangle$. The
laser is next switched off adiabatically while the photon is in
the cavity, such that the population is transferred by STIRAP from
the state $|g,1\rangle$ to the collective state
$|g^\prime,0\rangle$. In this final state the photon is stored,
shared among all the atoms.

A related adiabatic process is constructed here in order to transfer adiabatically the initial entangled state
$|g^\prime,0\rangle$ to a superposition of the marked states using an inverse
fractional stimulated Raman adiabatic passage (if-STIRAP). This
if-STIRAP is the time inversion of the fractional STIRAP
(f-STIRAP) which transfer population from a single state to a
superposition of states \cite{fstirap}. The if-STIRAP allows one
here to transfer the population from the superposition
$|g^\prime,0\rangle$ to the marked state.

\subsection{Model Hamiltonian}
The Hamiltonian describing the system of $N$ atoms is
\BE
 H_0=\delta \sum_{j=1}^M |g_j^\prime\rangle  \langle g_j^\prime| + \omega \sum_{j=1}^N |e_j\rangle \langle e_j| .
\EE
 We consider a cavity mode of frequency $\omega$ and coupling strength $G$ together with
 two lasers of frequencies $\omega$, $\omega-\delta$
 and  pulse shapes $\Omega(t)$, $\Omega^\prime(t)$ which do not grow with $N$.
 The resonant driving provided by the cavity-laser-atom system is described by
\BEA
 V&=&\omega a^\dagger a  + G a \sum_{j=1}^N |e_j\rangle  \langle g_j| \NN\\
 &+&  \left[\Omega e^{i \omega t} +\Omega^\prime e^{i (\omega-\delta) t}\right]
 \sum_{j=1}^{N}  |g^\prime_j \rangle \langle e_j| +\rm{h. c.}  \label{HD}
\EEA
The full Hamiltonian $H=H_0+V$ is block diagonal.
Each block may be labeled by the number $k$ of photons  which are present in
the cavity when the $N$ atoms are in their ground state
$|g\rangle$. The corresponding multipartite state $|g_1 \cdots
g_N  \rangle \otimes |k \rangle$ is denoted $|g,k\rangle$. All the
states which are coupled to $|g,k\rangle$ span a subspace whose
projection operator is ${\mathcal{P}}_k$. The Hamiltonian may thus be rewritten as $H=\sum_{k=0} {\mathcal{P}}_k H {\mathcal{P}}_k$.
We shall focus on the block ${\mathcal{P}}_1 H {\mathcal{P}}_1$ associated with a single photon in the
cavity.

The  multipartite state  $|g,1\rangle$ is coupled by the cavity to any state $
|g_1 \cdots  g_{j-1} e_{j} g_{j+1}   \cdots \rangle \otimes
|0\rangle \equiv |e_j,0\rangle $ for which the atom $j$  is in its excited state while the other atoms remain in their ground state $|g\rangle$.
Each state $|e_j,0\rangle$ is coupled by
the laser to the state $|g_1 \cdots  g_{j-1} g^\prime_{j}
g_{j+1}  \cdots g_N \rangle \otimes  |0\rangle \equiv |g^\prime_j,0\rangle$ where the atom $j$ is in the ground state $|g^\prime\rangle$ while the other atoms are in the ground state $|g\rangle$.
This coupling scheme is illustrated in Fig. 2 a, and the corresponding projection operator, spanning this subspace closer under $H$, reads
\BE
{\mathcal{P}}_1=\sum_{j=1}^N \left({\mathcal{P}}_{|g_j^\prime,0\rangle}+{\mathcal{P}}_{|e_j,0\rangle}\right)+{\mathcal{P}}_{|g,1\rangle}.
\EE

\begin{center}
\begin{figure}
\includegraphics[scale=0.55]{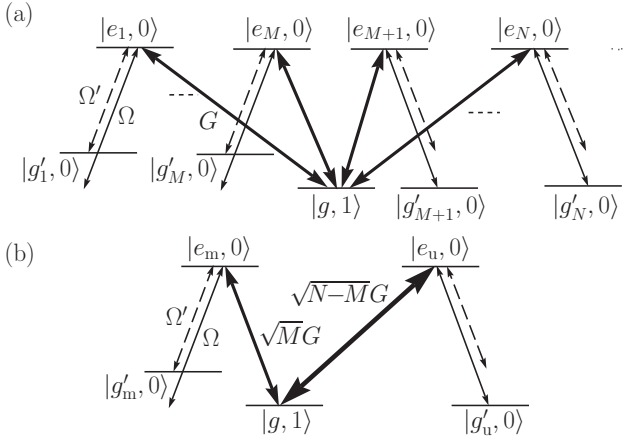}
\caption{(a) Coupling scheme in the cavity. The cavity $G$, laser
$\Omega$, laser $\Omega'$ Rabi frequency are featured by
respectively thick, thin, and dashed arrows. (b) Equivalent scheme
where the states $|g'_i,0\rangle$ and $|e_i,0\rangle$), $i=1,M$
(resp. $i=M+1,N$) of frame (a) form the collective marked ground
($|g'_{\rm m},0\rangle$) and excited ($|e_{\rm m},0\rangle$) states [resp. the
collective unmarked ground ($|g'_{\rm u},0\rangle$) and excited
($|e_{\rm u},0\rangle$) states]. The effective cavity Rabi frequency to
the collective marked (resp. unmarked) excited state is
$\sqrt{M}G$ (resp. $\sqrt{N-M}G$).} \label{FigStar}
\end{figure}
\end{center}

\subsection{Collective marked and unmarked dressed states}

In order to remove the oscillatory time dependence introduced by
diagonal terms, we consider atomic states which are dressed by laser and
cavity photons through the resonant transformation
\BEA
 R&=&e^{-i \delta t} \sum_{j=1}^M |g^\prime_j,0\rangle \langle g^\prime_j,0|+ \sum_{j=M+1}^N |g^\prime_j,0\rangle \langle g^\prime_j,0|\NN\\
 &+&e^{-i \omega t} |g,1\rangle \langle g,1|+e^{-i \omega t} \sum_{j=1}^{N} |e_j,0\rangle \langle e_j,0|.
\EEA
This results in the new Hamiltonian
\BE
H_R=R^{\dagger} ({\mathcal{P}}_1 H {\mathcal{P}}_1) R - i R^{\dagger} \dot R,
\EE
where the dot denotes time derivative.

Among the marked atoms, none plays a privileged role. Similarly, the unmarked atoms are all equivalent.
It is therefore appealing to treat them collectively.
We consider the uniform
superposition of marked (resp. unmarked) ground states \BEA
|g^\prime_{\rm m},0\rangle&=&\frac{1}{\sqrt{M}}\sum_{j=1}^{M}|g^\prime_j,0\rangle,\NN\\
|g^\prime_{\rm u},0\rangle&=&\frac{1}{\sqrt{N-M}}\sum_{j=M+1}^{N}|g^\prime_j,0\rangle,
\EEA
and the uniform superposition of excited states associated
with the marked (resp. unmarked) atoms
\BEA
|e_{\rm m},0\rangle&=&\frac{1}{\sqrt{M}}\sum_{j=1}^{M}|e_j,0\rangle.\NN\\
|e_{\rm u},0\rangle&=&\frac{1}{\sqrt{N-M}}\sum_{j=M+1}^{N}|e_j,0\rangle.
\EEA
These states, together with $|g,1\rangle$, define a subspace which is relevant for the Grover search.
Indeed, it can be shown that this subspace is closed under $H_R$,
\BE
H_R={\mathcal{P}} H_R {\mathcal{P}} +(I-{\mathcal{P}}) H_R ({\mathcal{P}}_1-{\mathcal{P}}),
\EE
where
\BE
{\mathcal{P}}={\mathcal{P}}_{|g^\prime_{\rm m},0\rangle}+{\mathcal{P}}_{ |g^\prime_{\rm u},0\rangle}+{\mathcal{P}}_{|e_{\rm m},0\rangle} +{\mathcal{P}}_{|e_{\rm u},0\rangle} +{\mathcal{P}}_{|g,1\rangle}.
\EE
We may thus restrict our attention to the Hamiltonian
\BEA
H_1&\equiv &{\mathcal{P}} H_R {\mathcal{P}} \NN\\
&=& G \left( \sqrt{M} |e_{\rm m},0\rangle + \sqrt{N-M} |e_{\rm u},0\rangle \right) \langle g,1|\NN \\
&+&\Sigma' |g'_{\rm m},0\rangle \langle e_{\rm m},0|+\Sigma |g'_{\rm u},0\rangle \langle e_{\rm u},0| + {\rm h.c.},\label{H1}
\EEA
where $\Sigma=\Omega+e^{-i \delta t} \Omega^\prime$ and $\Sigma^\prime=\Omega^\prime+e^{i \delta t} \Omega$. This Hamiltonian is schematically represented in Fig. 2 b.

\subsection{Effective Hamiltonian}
The Hamiltonian $H_1$ can be rewritten in a way which is prone to a further reduction. Indeed, notice that the first term of (\ref{H1}) actually features a linear combination of $|e_{\rm m},0\rangle$ and $|e_{\rm u},0\rangle$ which is nothing but the uniform superposition over all atoms $\frac{1}{\sqrt{N}}\sum_{j=1}^{N}|e_j,0\rangle \equiv |e,0\rangle$.
This is expected since the state $|g,1\rangle$ is coupled indifferently to any of the state $|e_j,0\rangle$, marked or not, as is seen for instance in Fig. 2 a.
By contrast, owing to the detuning, the $g'-e$ transitions are not equivalent in the marked and unmarked cases.   Hence, the last two terms of (\ref{H1}) involve either $|e_{\rm m},0\rangle$ or $|e_{\rm u},0\rangle$.
We may express $H_1$ in terms of $|e,0\rangle$ and a state $|e_{\perp},0\rangle$ which is orthogonal to the uniform superposition,
\BEA
|e,0\rangle &=& \sqrt{f} |e_{\rm m},0\rangle +\sqrt{1-f} |e_{\rm u},0\rangle, \qquad f\equiv \frac{M}{N}   \NN\\
|e_{\perp},0\rangle &=& \sqrt{1-f} |e_{\rm m},0\rangle-\sqrt{f} |e_{\rm u},0\rangle.
\EEA
The result reads
\BE H_1= H_{e_{\perp}}+H_e+H_{e_{\perp} e}+H_{e e_{\perp}}
\label{H1block}, \EE where $H_e$ and $H_{e_{\perp}}$ are coupled
to each other only through $H_{e_{\perp}e}$ and $H_{e
e_{\perp}}=H_{e_{\perp}e}^\dagger$,

\BEA
H_{e_{\perp}}\!\!\!&=&\!\!\!\sqrt{1-f} \Sigma'  \ |g'_{\rm m},0\rangle \langle e_{\perp},0|-\sqrt{f} \Sigma \ |g'_{\rm u},0\rangle \langle e_{\perp},0| + {\rm h.c.} \NN\\
H_e\!\!\!&=&\!\!\!  \sqrt{N}  G \  |e,0\rangle \langle g,1| + {\rm h.c.} \NN\\
H_{e_{\perp} e}\!\!\!&=&\!\!\!\sqrt{f} \Sigma'  \ |g'_{\rm m},0\rangle \langle e,0|+\sqrt{1-f} \Sigma \ |g'_{\rm u},0\rangle \langle e,0|.\label{H1e}
\EEA

\begin{center}
\begin{figure}
\includegraphics[scale=0.6]{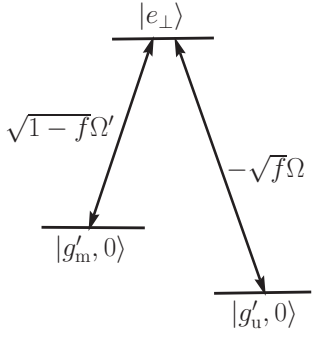}
\caption{Coupling scheme in the $\Lambda$ system corresponding to
the effective Hamiltonian (\ref{Heff}) with a fraction $f$($=M/N)$ of solution to the search problem.} \label{FigHeff}
\end{figure}
\end{center}

As just mentioned, it is the Hamiltonian  $H_{e_{\perp}}$ which
discriminates the marked and unmarked states whereas  $H_{e}$ only
features uniform superpositions over all the atoms. The subspace
of interest is thus spanned by the projector \BE
{\mathcal{P}}_{e_{\perp}}={\mathcal{P}}_{|g^\prime_{\rm m},0\rangle}+{\mathcal{P}}_{ |g^\prime_{\rm
u},0\rangle}+{\mathcal{P}}_{|e_\perp\rangle}. \EE With the help of a unitary
transformation $T=\exp (W)$ we can perturbatively block diagonalize
the Hamiltonian $H_1$. In this relevant subspace, the first
correction to $H_{e_{\perp}}$ is obtained by the partitioning
technique, which is recalled in appendix, as
$\frac{1}{2}{\mathcal{P}}_{e_{\perp}} \left[H_{e e_{\perp}}, \ep W \right]
{\mathcal{P}}_{e_{\perp}}$. This quantity, featuring denominators which are
the difference of the eigenvalues of  $H_{e_{\perp}}$ and $H_{e}$,
decreases at least as $1/ N G^2$ (with additional decreasing contributions due to $f$).
As a consequence, we shall
neglect this correction together with the nonresonant components
in $\Sigma$ and $\Sigma'$.
We therefore consider  the effective Hamiltonian
\BE
H_{\rm eff} = \sqrt{1-f} \Omega'  \ |g'_{\rm m},0\rangle
\langle e_{\perp},0|-\sqrt{f} \Omega \ |g'_{\rm u},0\rangle
\langle e_{\perp},0| + {\rm h.c.} ,\label{Heff}
\EE
which is schematically depicted in
Fig. \ref{FigHeff}.

\section{Grover search}
\subsection{Starting and final points}
Our aim is to transfer adiabatically the population from the
initial state $|g^\prime,0\rangle$,
\BE
 |g',0\rangle \equiv\frac{1}{\sqrt{N}}\sum_{j=1}^{N}|g'_j,0\rangle ,
\EE
which gives no privileged role to any of the $N$
states $|g^\prime_j,0\rangle$ to a final state  which coincides
with the collective marked state $|g^\prime_{\rm m},0\rangle$ in a time which
scales as $\sqrt{1/f}$ where $f=M/N$ is the fraction of solutions. The population transfer mechanism is most
easily revealed in the basis of the instantaneous eigenstates  of
$H_{{\rm eff}}(t)$
\BEA
|0\rangle(t) &=& \cos \theta(t) \ |g^\prime_{\rm m},0\rangle
-\sin \theta(t) \ |g^\prime_{\rm u},0\rangle\label{adiab}\\
|\pm \Lambda\rangle(t) &=& \frac{1}{\sqrt{2}}\left(\sin \theta(t)
\ |g^\prime_{\rm m},0\rangle +\cos \theta(t) \ |g^\prime_{\rm
u},0\rangle \pm |e_\perp,0 \rangle \right),\NN \EEA pertaining to
the eigenvalues 0 and $\pm \Lambda(t)$ where \BEA
\Lambda(t)=\sqrt{(1-f){\Omega^\prime}^2(t)+f\Omega^2(t)}.
\label{Lambda} \EEA Note that $|0\rangle$ has no component on the
collective excited states $|e_\perp,0 \rangle$  and is therefore a
so-called dark state which is immune to loss by spontaneous
emission (in contrast to the states $|\pm \Lambda\rangle$). The
instantaneous angle $\theta(t)$ is defined through the relation
\BEA \tan \theta(t) &=& -\sqrt{\frac{1-f}{f}}
\frac{\Omega^\prime(t)}{\Omega(t)}. \label{tan} \EEA Requiring the
instantaneous eigenstate (\ref{adiab}) to coincide at the initial
time $t_{\rm i}$ with  the uniform superposition
\begin{equation}
\label{superp} |g^\prime,0\rangle={\sqrt{f}}  |g^\prime_{\rm m},0
\rangle + \sqrt{1-f}  |g^\prime_{\rm u},0 \rangle
\end{equation}
 and at the final time with
the collective marked state $|g^\prime_{\rm m},0\rangle$ entails
that \BE \label{tanif} \tan \theta(t_{\rm i})=
-\sqrt{\frac{1-f}{f}},\quad \tan \theta(t_{\rm f})= 0 . \EE This
implies that the two pulses must be switched on simultaneously,
$\Omega^\prime(t_{\rm i})=\Omega(t_{\rm i})$, and that the pulse
$\Omega^\prime$ is to be turned off before $\Omega$. This process
can be named inverse fractional STIRAP (if-STIRAP) since it allows
the transfer by a STIRAP type process from a superposition of
states to a single state \cite{STIRAP}. In the adiabatic representation
(\ref{adiab}), the effective Hamiltonian reads \BEA H_{{\rm
eff}}^{\rm ad}&=&
\Lambda \left(|+\Lambda\rangle \langle +\Lambda| \ + \ |-\Lambda\rangle \langle -\Lambda|\right)\NN\\
&+& \frac{i \dot \theta}{\sqrt{2}}  \left(|+\Lambda\rangle \langle 0|+|-\Lambda\rangle \langle 0|-{\rm h.c.}\right) \label{Had}
\EEA
where  $\dot \theta =\frac{1}{1+\tan^2 \theta}
\frac{d}{dt} \tan \theta$. In the adiabatic regime, the
transitions between instantaneous eigenstates are negligible. This
will be achieved if the Hamiltonian varies sufficiently slowly in
time so as to keep $\dot \theta \ll \Lambda$. On the other hand,
we wish to control the process duration and, in particular, to
prevent it from becoming arbitrary large.
This can be achieved in several ways, two of which shall be considered explicitly.

\subsection{Local adiabatic conditions}
In order to control the process duration, as
proposed in \cite{roland}, we choose to require $\dot \theta$ and
$\Lambda$ to be in a constant (small) ratio $\ep$ at all times,
independently of $N$: \BE \label{epLambda} \dot \theta = \ep
\Lambda. \EE \
Given a laser pulse $\Omega$, this equation will allow us to
determine the pulse $\Omega^\prime$ which is needed to remain in
the instantaneous eigenstate $|0\rangle(t)$ with a significant probability
throughout the process, starting from the
uniform superposition $|0\rangle(t_{{\rm i}})=|g^\prime,0\rangle$,  and
ending up in the marked state $|0\rangle(t_{{\rm
f}})=|g^\prime_{{\rm m}},0\rangle$ after some time
$t_{{\rm f}}-t_{{\rm i}}$ which achieves the optimal scaling with $N$.

Indeed, let us first rewrite $\Lambda$ defined in
(\ref{Lambda}) in terms of $\tan
\theta$ through (\ref{tan}) as
\BE
\Lambda=\sqrt{f}\sqrt{1+\tan^2 \theta}  \Omega .
\EE
We then obtain from (\ref{epLambda}) a differential equation for $\tan
\theta$, {\em i. e.}, for the ratio $\Omega^\prime/\Omega$
\BE
\frac{d  \tan \theta}{\left(1+\tan^2 \theta\right)^{3/2}}= \ep \sqrt{f} \Omega dt.
\EE
Its solution satisfying the initial condition (\ref{tanif}) reads \BE
\frac{\Omega^\prime(t)}{\Omega(t)}=\frac{1-\ep {\mathcal
A}(t)\sqrt{\frac{f}{1-f}}}{\sqrt{1+\ep {\mathcal
A}(t)\{2\sqrt{\frac{1-f}{f}}-\ep {\mathcal A}(t)\}}} , \label{rho}
\EE where we define ${\mathcal A}(t)\equiv\int_{t_{{\rm i}}}^t du
\Omega(u)$.
Upon specifying that at time $t_{{\rm f}}$ the ratio of the pulses vanishes, one deduces from  (\ref{rho}) that  $\ep {\mathcal
A}(t_{{\rm f}})=\sqrt{(1-f)/f}$. Expressing the total area of the
pulse $\Omega$ in terms of its peak amplitude $ {\Omega}_{0}$ and the process duration $\mathcal{T}$,
${\mathcal A}(t_{{\rm f}})={\Omega}_{0} \mathcal{T}$, we finally obtain
\BE \label{area}
{\Omega}_{0}\mathcal{T}=\frac{1}{\ep}\sqrt{\frac{1-f}{f}}.
 \EE
This expression shows that  the search duration scales as $f^{-1/2}$ for a peak amplitude which is independent
of $f=M/N$.
Equivalently, we can increase the peak amplitude as $f^{-1/2}$ for a
constant duration $\mathcal{T}$.

We now determine the population which can be reached by adiabatic
passage. With the choice (\ref{epLambda}), we see from (\ref{Had}) that $\Lambda$ appears as factor in the Hamiltonian $H_{{\rm eff}}^{\rm ad}$.
Hence, we can define a new time
$\tau(t)=\int_{t_{{\rm i}}}^t ds \Lambda(s)$ so that the
Hamiltonian $H_{{\rm eff}}^{\rm ad} / \Lambda$ is time
independent. It follows that the survival probability amplitude of
the state $|0 (t)\rangle$ is \BE \langle 0 (t)| \exp{\left(-i
\tau(t) \frac{ H_{{\rm eff}}^{\rm ad}}{\Lambda}\right)}|0 (t_{{\rm
i}}) \rangle =\frac{1+\ep^2 \cos\lambda\tau(t) }{1+\ep^2},
\label{pop0} \EE where $\lambda\equiv \sqrt{1+\ep^2}$. The system
therefore stays in the state $|0 (t)\rangle$ with a probability
satisfying at all times
\begin{equation}
\label{P0}
P_0(t)\ge\left(\frac{1-\ep^2}{1+\ep^2}\right)^2\sim
1-4\ep^2.
\end{equation}
The non-adiabatic losses (to the states $|\pm \Lambda\rangle$ which contain some component on the excited state $|e_\perp\rangle$) are therefore never larger than $4 \ep^2$. Furthermore, since the states $|0\rangle$ and $|g'_{{\rm m}},0\rangle$ coincide at the final time, (\ref{P0}) also implies that the collective marked state ends up with a population of the order of $1-4\ep^2$.

Returning to the diabatic representation, in which the state vector is denoted  $|\phi (t)\rangle$ , one can determine the full population dynamics of the collective states featured in the effective Hamiltonian (\ref{Heff}).
The population of the collective marked state is found to be
\BEA
P_{{\rm m}}(t)& \equiv&|\langle g'_{{\rm m}},0| \phi (t) \rangle|^2 \NN\\
&=&|\frac{\ep}{\lambda}
\sin \lambda\tau(t) \sin \theta(t) +\frac{1+\ep^2 \cos
\lambda\tau(t) }{1+\ep^2} \cos \theta(t)|^2, \NN\\ \label{Pm}
\EEA
 where
$\theta(t)$ is given by (\ref{tan}) and (\ref{rho}).
Recalling from (\ref{tanif}) that, at the final time, $\sin \theta(t_{{\rm f}})=0$, one deduces that
\BE
P_{{\rm m}}(t_{{\rm f}})=
\left(\frac{1+\ep^2 \cos \lambda\tau(t_{{\rm f}}) }{1+\ep^2}\right)^2 \ge \left(\frac{1-\ep^2}{1+\ep^2}\right)^2. \label{Pmf}
\EE
The population $P_{{\rm u}}(t)\equiv|\langle g'_{{\rm u}},0| \phi (t) \rangle|^2$ of the collective unmarked state is given by an expression similar  to (\ref{Pm}) where $\sin \theta(t)$ and $\cos \theta(t)$ are interchanged. In particular, this entails that, at the final time, it is of order $\ep^2$,
\BE
P_{{\rm u}}(t_{{\rm f}})=\frac{\ep^2}{1+\ep^2}
\sin^2 \lambda\tau(t_{{\rm f}}).
\EE
Finally, the population of the collective excited state $|e_\perp\rangle$ is
\BE
P_{{e_\perp}}(t)=\frac{2\ep^2}{(1+\ep^2)^2}(1-\cos \lambda\tau(t))^2, \label{Pe}
\EE
which puts a bound on the decoherence present in the system since the population of this excited state is never larger than $\ep^2$.

\subsection{Tailored Gaussian pulses}
The local adiabatic conditions considered in the preceding section
allowed us to determine explicitly the scaling of the search
duration and the population of the collective marked state. This
is only a particular choice.
The proposed scheme is
robust in this respect.
Indeed, as long as the initial
and final conditions (\ref{tanif}) are satisfied and one goes
adiabatically from one to the other, the desired transfer of
population can be achieved in a time which scales appropriately. In practical applications, it may be
easier to consider Gaussian pulses possibly with plateaus.

In order to satisfy the initial condition, the two pulses $\Omega$ and $\Omega'$ that are used in the current scheme are to be turned on simultaneously. The pulse
$\Omega'$ is a standard Gaussian pulse $\Omega'=\Omega_0e^{-(t/T)^2}$. As for $\Omega$, it has a Gaussian switching on, and when the peak is reached,
the pulse is kept at this value for a time $\alpha T$ before being switched off according to a Gaussian,
\BEA
\Omega=
\left\{   \begin{array}{ll}
      \Omega_0e^{-(t/T)^2} & t \geq 0 \\
       \Omega_0 & 0 \leq t \leq \alpha T \\
        \Omega_0e^{-(t/T-\alpha)^2} & t \geq \alpha T
   \end{array}
   \right. . \label{tailor}
\EEA
We show numerically in the next section that with such pulses the search duration has a scaling which is close to optimal for $\alpha \gtrsim 1.5$.

\section{Discussion}
The current adiabatic search scheme is robust in the sense that it
leaves some flexibility on the pulses that are used. We presented
explicitly two choices of pulses which will now be illustrated and
compared.
\begin{center}
\begin{figure}
\includegraphics[scale=0.7]{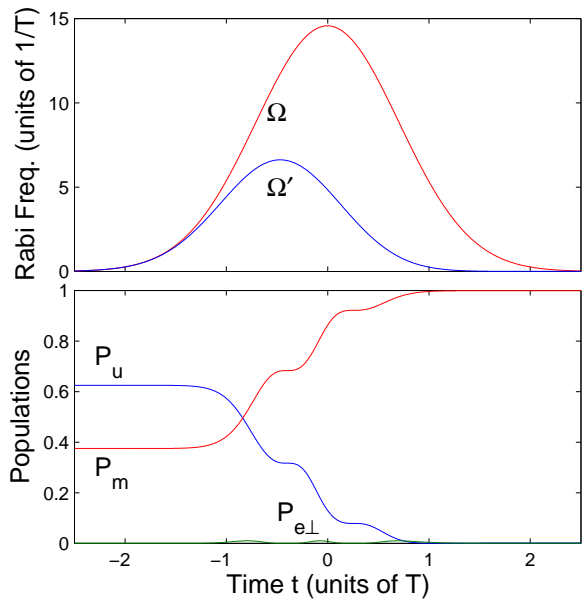}
\caption{Analytical dynamics for the effective Hamiltonian
(\ref{Heff}) with $f=M/N=3/8$, $\varepsilon=0.05$ and Rabi frequencies given by a Gaussian profile $\Omega(t)=\Omega_{0}e^{-(t/T)^2}$  and expression (\ref{rho}) for $\Omega^{\prime}(t)$.
Here
$\Omega_{0}T=\sqrt{(1-f)/f}/\varepsilon\sqrt{\pi}$  in order to
satisfy the conditions (\ref{tanif}) and (\ref{epLambda}).
The final population transfer to the collective marked state is larger than 0.99.
Top: Rabi frequencies. Bottom:
Populations of the collective marked  ($P_{\rm
m}$), unmarked ($P_{\rm u}$) and excited ($P_{e\perp}$) states as a function of dimensionless time.} \label{Dynamics}
\end{figure}
\end{center}

We first consider the case of  (\ref{epLambda}) which amounts to
requiring local adiabatic conditions. The pulses are determined by
(\ref{rho}) and displayed in the upper panel of Fig.
\ref{Dynamics} for $f=M/N=3/8$, $\varepsilon=0.05$,
and a Gaussian pulse $\Omega$ of characteristic duration $T$.
This value of $\varepsilon$  implies
that the lower bound (\ref{P0}) of the probability $P_0$ to remain in the instantaneous eigenstate is 0.99.
The
population dynamics of the collective marked, unmarked and excited states is given by the analytical expressions
(\ref{Pm})-(\ref{Pe}) and is depicted in the lower panel. As
predicted by (\ref{Pmf}), the transfer to the marked state is very
efficient with a low transient population in the excited states
stemming from the fact that the dynamics remains in the
instantaneous decoherence-free eigenstate $|0\rangle(t)$ in the
adiabatic limit. We remark that the choice (\ref{epLambda}), which
leads to the seemingly complicated pulse relation (\ref{rho}),
gives in practice a simple smooth bell-shaped pulse (see upper panel of Fig.
\ref{Dynamics}). Increasing $f$ (i. e. increasing $M$ for a given
$N$) allows the reduction of the pulse area with the same
efficiency, as shown by (\ref{area}).

\begin{center}
\begin{figure}
\includegraphics[scale=0.7]{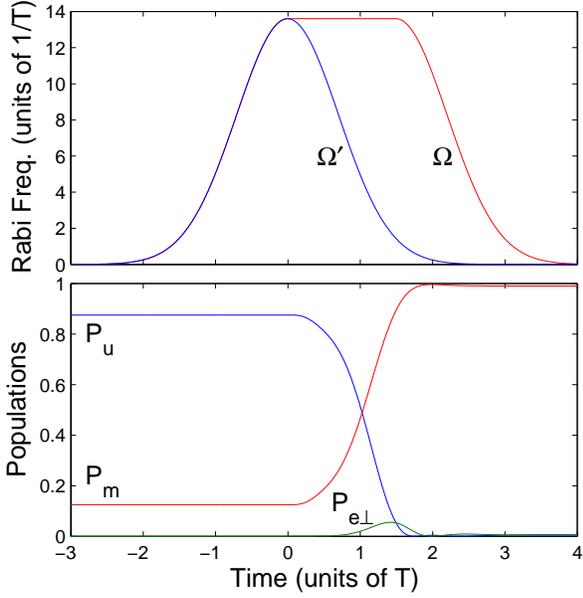}
\caption{Numerical dynamics for the effective Hamiltonian
(\ref{Heff}) with $f=M/N=1/8$ and Rabi frequencies chosen as $\Omega'(t)=\Omega_{0}e^{-(t/T)^2}$ and a Gaussian profile with a plateau for $\Omega(t)$ as in (\ref{tailor})  in order to satisfy (\ref{tanif}).
Here $\Omega_{0}T$ is such that the final population transfer to the marked state is 0.99. Top: Rabi frequencies. Bottom: Populations of the collective marked  ($P_{\rm
m}$), unmarked ($P_{\rm u}$) and excited ($P_{e\perp}$) states as a function of dimensionless time.} \label{Dynamics_PG_M_1}
\end{figure}
\end{center}

\

\begin{center}
\begin{figure}
\includegraphics[scale=0.7]{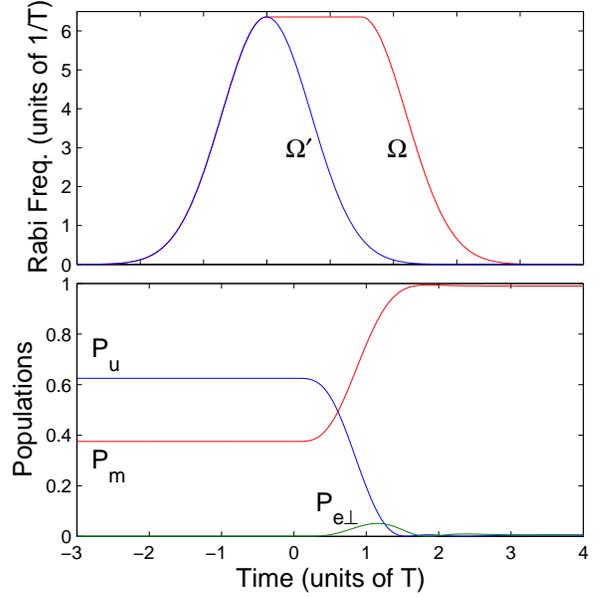}
\caption{Same as Fig. \ref{Dynamics_PG_M_1} but with $f=M/N=3/8$.}
\label{Dynamics_PG_M_3}
\end{figure}
\end{center}
\begin{center}
\begin{figure}
\includegraphics[scale=0.7]{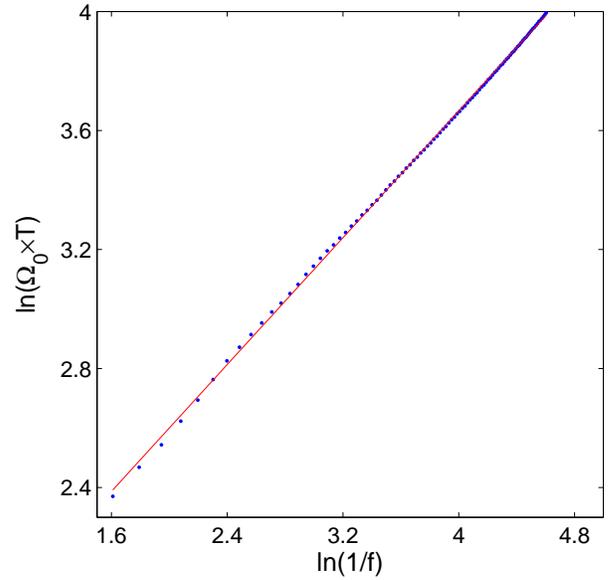}
\caption{Scaling of $\Omega_0 T$, with $T$ the process duration, as a function of the fraction $f$ of solution for the search
with tailored Gaussian pulses as in Fig. \ref{Dynamics_PG_M_1}. The plateau duration is fixed to 1.5 $T$ and the final population of the collective marked state is 0.99. The natural logarithm of $\Omega_0 T$ as a function of
the natural logarithm of $1/f$ obtained numerically (dotted line) is well fitted by the straight line of slope 0.53.}
\label{Scaling}
\end{figure}
\end{center}

We now turn to the case of tailored Gaussian pulses.
Figures \ref{Dynamics_PG_M_1} and \ref{Dynamics_PG_M_3} show the
dynamics for such pulses with $f=M/N=1/8$ and $f=3/8$
respectively.  We have here chosen a plateau of duration $1.5 T$ ($\alpha=1.5$).
The pulse area of $\Omega'$ has been chosen such
that the population transfer to the marked state is approximately
0.99, i.e. as in the conditions of Fig. \ref{Dynamics}. Figures
\ref{Dynamics_PG_M_1} and \ref{Dynamics_PG_M_3} show similar
features as \ref{Dynamics}: low transient population in the
excited states, comparable pulse area for the same efficiency,
reduction of the pulse area with the same efficiency for
increasing $M$. We remark however two different properties: (i)
For a similar efficiency, there is a larger transient population
in the upper state for the cases of Figs. \ref{Dynamics_PG_M_1}
and \ref{Dynamics_PG_M_3}; (ii) The population dynamics of the
latter cases do not exhibit the oscillations noticed in Fig.
\ref{Dynamics}. Both features can be interpreted using
superadiabatic basis that are better adapted to describe the
dynamics \cite{Berry,Joye}. This superadiabatic transport allows
one to explain the final revival of the transient ``lost''
population in the upper state shown in Figs. \ref{Dynamics_PG_M_1}
and \ref{Dynamics_PG_M_3} \cite{Elk,Drese}. Furthermore, the use
of analytic pulses in the cases of Figs. \ref{Dynamics_PG_M_1} and
\ref{Dynamics_PG_M_3} explain the non-oscillatory transfer. For
the preceding case, the local adiabatic control (\ref{epLambda})
induces the condition (\ref{rho}) which prevents in general the
analyticity of one of the pulse (here $\Omega'$ if $\Omega$ is
chosen analytic). In this non-analytic case, through non-adiabatic
transitions in a superadiabatic basis (of order related to the
order of the discontinuous derivative), population transfer occurs
in an oscillatory manner as shown in Fig. \ref{Dynamics}.

Figure \ref{Scaling} displays the scaling of the search duration $T$  for the scheme
which uses the pulses of the the form (\ref{tailor}) as shown in the upper panel of Fig. \ref{Dynamics_PG_M_1}.
The duration of the plateau has been kept to $1.5 T$ ($\alpha=1.5$)
and the quantity $\Omega_0T$ has been determined numerically from the requirement that
the population transfer to the collective marked state reaches 0.99. We
obtain a duration search scaling essentially as $f^{-0.53}$, which
demonstrates the approximate quadratic speedup of the search.

\begin{center}
\begin{figure}
\includegraphics[scale=0.7]{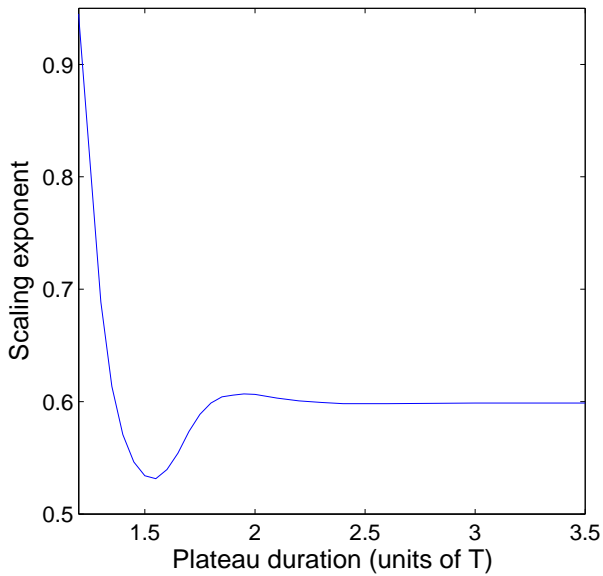}
\caption{Power $\beta$ of the scaling $f^{-\beta}$ of $\Omega_0 T$ as a function of the plateau durations $\alpha T$  for tailored Gaussian pulses.
The scaling exponent $\beta$ is obtained as in Fig. \ref{Scaling} (see text for details). }
\label{power}
\end{figure}
\end{center}

Finally, we investigate the role of the plateau in the efficiency of the search.
The scaling exponent  $\beta$ of the search duration for various
plateau durations $\alpha T$ is analyzed in Fig. \ref{power}. For each value
of $\alpha$, the power $\beta$ is determined numerically by a
linear fit of the log-log representation of $\Omega_0T$ as a
function of $1/f$ (obtained in the same conditions as for Fig.
\ref{Scaling}). Figure \ref{power} shows that the power $\beta$
decreases to 0.53 as the dynamics becomes more adiabatic until
$\alpha\approx 1.5$ before increasing slightly and saturating to 0.6 for $\alpha>2.5$.

Before concluding, let us make some remarks concerning the
experimental implementation of this adiabatic Grover search.
Firstly, the scheme described here requires to trap atoms in a
cavity, a task which can be achieved for instance using a
standing-wave dipole-force trap \cite{Kuhn}. As a realistic
$\Lambda$ atomic scheme, we can consider the typical
2$^3S_1-2^3P_0$ transition in metastable helium which is of
linewidth $\Gamma\sim10^7$ s$^{-1}$ and Rabi frequency
$\Omega\sim10^8 \sqrt{I}$ s$^{-1}$ (with the intensity $I$ in
W$/$cm$^2$). In order to neglect spontaneous emission, as assumed
in (\ref{HD}), we require the condition $(\Omega_{0}T)^2\gg\Gamma
T$. It is fulfilled when $\Omega_{0}T\gg 1$ and
$\Omega_{0}\gg\Gamma $, which are well satisfied in practice,
{\emph e.g.}, for $I_{0}\sim10^4$ W/cm$^2$ and $T\sim10$ ns.
Finally, the feasibility of fixing the ratio of two pulses (here
essentially required at early times in the local adiabatic
conditions approach) has been shown in \cite{fstirap_exp} using
acousto-optical modulation of a cw laser in such a nanosecond
regime. This step is not required in the other approach using
Gaussian pulses with plateaus. One can mark the atoms by an ac
Stark shift using a magnetic field, for instance from a focused
laser beam for spatial resolution.

\section{Conclusions}

We have investigated theoretically a physical implementation of
the adiabatic Grover search using a cavity-laser-atom system and
robust processes related to STIRAP. The calculation has been
conducted with pulses based on the constraint (\ref{epLambda})
that has allowed us to prove the scaling analytically. We have
checked the robustness of the $f^{-1/2}$ scaling by numerical
simulations using other less restrictive adiabatic pulse shapes
satisfying (\ref{tanif}). The case of Gaussian pulses with
plateaus, which are routinely used in laboratories, has been
studied. In particular,  a scaling close to optimal is obtained
when, after a Gaussian switching on, one of the two pulses is kept
constant for a time equal to (or larger than) 3/2 of its
characteristic time. Finally, the experimental implementation of
the presented scheme has been briefly discussed.


\section*{Acknowledgments}
The authors are grateful to N. J. Cerf and H.-R. Jauslin for
useful discussions and acknowledge the support from the EU
projects COVAQIAL and QAP, from the Belgian government programme
IUAP under grant V-18, and from the Conseil R\'egional de
Bourgogne. S.G. acknowledge support from the French Agence
Nationale de la Recherche (ANR CoMoC).


\begin{appendix}
\section{Partitioning}

Let us consider a system whose Hilbert space is partitioned into two orthogonal Hilbert spaces by means of the projection operators ${\mathcal{P}}_A$ and ${\mathcal{P}}_B=1-{\mathcal{P}}_A$.
The pertaining Hamiltonian can be written as
\BE
H=H_A+H_B+H_{AB}+H_{BA} \equiv H_A+H_B+ \epsilon  V,
\EE
where $\epsilon$ is a formal small (or more precisely ordering) parameter.
The purpose is to achieve a perturbative block-diagonalization of $H$.
The leading contribution from the off-diagonal blocks to the diagonal ones can be extracted by the partitioning technique \cite{GJ}.
There is a unitary transformation $\exp (\epsilon W)$ such that
\BE
e^{-\epsilon W} H e^{\epsilon W}=H_A+H_B+\frac{1}{2}[\epsilon V, \epsilon W] +\epsilon^3 V', \label{corr2}
\EE
with the property that $[\epsilon V, \epsilon W]={\mathcal{P}}_A[\epsilon V, \epsilon W]{\mathcal{P}}_A+{\mathcal{P}}_B[\epsilon V, \epsilon W]{\mathcal{P}}_B$.
The leading correction to the diagonal blocks is thus of order 2 while the remainder ($\epsilon^3 V'$) is of order 3.
It has been shown that $\epsilon W = \epsilon W_{BA}- \epsilon W_{BA}^{\dagger}$ with $\epsilon W_{BA}$ given by
\BE
\epsilon W_{BA}=-\sum_{j,k}\frac{\left|\lambda^B_j\right\rangle \left\langle  \lambda^B_j\right| H_{BA}\left|\lambda^A_k\right\rangle\left\langle  \lambda^A_k\right|}{\lambda^B_j-\lambda^A_k}, \label{WBA}
\EE
where $\left|\lambda^A_k\right\rangle$ denotes the eigenvector of $H_A$ corresponding to the eigenvalue $\lambda^A_k$ (and similarly for $H_B$).
Recall that $\epsilon$ is only formally a small parameter (it can be set to 1). However, note from (\ref{WBA}) that $\epsilon W_{BA}$ and the high order terms in (\ref{corr2}) are smaller, the larger the differences in the eigenvalues of $H_A$ and $H_B$.

We now apply these results to the Hamiltonian (\ref{H1block}), taking $H_1$ for $H$,
${\mathcal{P}}_{e_{\perp}}$ for ${\mathcal{P}}_A$ and ${\mathcal{P}}_{e}$ for ${\mathcal{P}}_B$.
The dominant contribution from the off-diagonal blocks $H_{e_{\perp} e}$ and $H_{ ee_{\perp}}$ to $H_{e_{\perp}}$
is obtained from (\ref{corr2}),
\BEA
H_{e_{\perp}}^{(2)}&=&H_{e_{\perp}}+\frac{1}{2}{\mathcal{P}}_{e_{\perp}}[\epsilon V,\epsilon W ] {\mathcal{P}}_{e_{\perp}}\NN\\
&=&H_{e_{\perp}}+\frac{1}{2} \left(H_{e_{\perp} e} \epsilon W_{e e_{\perp}}  + {\rm h.c.}\right), \label{H2}
\EEA
where $\epsilon W_{e e_{\perp}}$ is given by (\ref{WBA}).
The eigenvalues of $H_{e}$ are  $\pm \sqrt{N} G$ and the corresponding eigenvectors read
\BEA
|\pm \sqrt{N} G\rangle&\equiv &\frac{1}{\sqrt{2}} \left(|e,0 \rangle \pm |g,1 \rangle \right).
\EEA
In the resonant approximation ($\Sigma\equiv\Omega+e^{-i \delta t} \Omega^\prime \approx \Omega$ and $\Sigma^\prime\equiv\Omega^\prime+e^{i \delta t} \Omega\approx \Omega^\prime$), the eigenvectors of $H_{e_{\perp}}$ are
\BEA
|0\rangle &=& \cos \theta \ |g^\prime_{\rm m},0\rangle
-\sin \theta \ |g^\prime_{\rm u},0\rangle\\
|\pm \Lambda\rangle &=& \frac{1}{\sqrt{2}}\left(\sin \theta \
|g^\prime_{\rm m},0\rangle +\cos \theta \ |g^\prime_{\rm u},0\rangle \pm
|e_\perp,0 \rangle \right),\NN
\EEA
with the instantaneous angle $\theta$ defined through $\tan \theta \equiv -\sqrt{(1-f)/f} \Omega^\prime/\Omega$ and
\BE
\Lambda\equiv \sqrt{(1-f){\Omega^\prime}^2+f\Omega^2}. \label{Lambda_app}
\EE
To determine the leading correction to $H_{e_{\perp}}$ given in (\ref{H2}), we note that $H_{e_{\perp} e} =H_{e_{\perp} e} {\mathcal{P}}_e =H_{e_{\perp} e}\left|e,0\right\rangle  \left\langle e,0\right| $.
Hence, we simply have to compute $\left\langle e,0\right|\epsilon W_{e e_{\perp}}$, which after some algebra, reads
\BEA
\left\langle e,0\right|\epsilon W_{e e_{\perp}}&=&\frac{\Lambda\left(\sqrt{f} \Omega' \sin \theta +\sqrt{1-f} \Omega \cos \theta\right)}{NG^2-\Lambda^2}\left\langle e_{\perp},0\right|\NN\\
&=&\frac{ \Lambda \sqrt{1-f}  (\Omega^2-{\Omega'}^2)\cos \theta}{(NG^2-\Lambda^2)\Omega}\left\langle e_{\perp},0\right|\NN
\EEA
The correction featured in (\ref{H2}) is then
\BEA
H_{e_{\perp} e} \epsilon W_{e e_{\perp}}&=& \frac{ \Lambda \sqrt{1-f}  (\Omega^2-{\Omega'}^2)\cos \theta}{(NG^2-\Lambda^2)\Omega}\NN\\
\times \left(\sqrt{f} \Omega'  \ |g'_{\rm m},0\rangle \right.&+& \left.\sqrt{1-f} \Omega \ |g'_{\rm u},0\rangle\right)
\left\langle e_{\perp},0\right| .
\label{corr}
\EEA
Recall that, by definition of the problem, the constraint imposed in this Grover search is that the pulse envelopes $\Omega$ and $\Omega'$ do not increase with $N$, i. e., are at most of order $N^0$.
It then follows from (\ref{Lambda_app}) that $\Lambda$ is at most of order $N^0$.
Note that the additional dependency on $N$ introduced through $f$ may only decrease $\Lambda$.
As a consequence, the correction (\ref{corr}) to $H_{e_{\perp}}$ decreases at least as $1/NG^2$ and may be discarded, giving rise to the effective Hamiltonian $H_{\rm eff} =H_{e_{\perp}}$ defined in (\ref{Heff}).
\end{appendix}


\begin{thebibliography}{99}
\bibitem{grover} L. K. Grover, Phys. Rev. Lett, {\bf 79}, 325 (1997).
\bibitem{nmr1} I. L. Chuang, N. Gershenfeld, and M. Kubinec, Phys. Rev. Lett. {\bf 80}, 3408 (1998).
\bibitem{nmr2} J. A. Jones, M. Mosca, and R. H. Hansen, Nature (London) {\bf 393}, 344 (1998).
\bibitem{nmr3} M. S. Anwar, D. Blazina, H. A. Carteret, S. B. Duckett, and J. A. Jones, Chem. Phys. Lett. {\bf 400}, 94 (2004).
\bibitem{optics1} P. G. Kwiat, J. R. Mitchell, P. D. D. Schwindt, and A. G. White, J. Mod. Opt. {\bf 47}, 257 (2000).
\bibitem{optics2} P. Walther, K. J. Resch, T. Rudolph, E. Schenck, H. Weinfurter, V. Vedral, M. Aspelmeyer, and A. Zeilinger, Nature (London) {\bf 434}, 169 (2005).
\bibitem{ions} K.-A. Brickman, P. C. Haljan, P. J. Lee, M. Acton, L. Deslauriers, and C. Monroe, Phys. Rev. A {\bf 72}, 050306(R) (2005).
\bibitem{qed1} F. Yamaguchi, P. Milman, M. Brune, J. M. Raimond, and S. Haroche, Phys. Rev. A {\bf 66}, 010302(R) (2002).
\bibitem{qed2} Z. J. Deng, M. Feng, and K. L. Gao, Phys. Rev. A {\bf 72}, 034306 (2005).
\bibitem{farhi} E. Farhi and S. Gutmann, Phys. Rev. A \textbf{57}, 2403 (1998).
\bibitem{nmr4} V. L. Ermakov and B. M. Fung, Phys. Rev. A {\bf 66}, 042310 (2002).
\bibitem{adiabatic1} E. Farhi, J. Goldstone, S. Gutmann, and M. Sipser,
e-print quant-ph/0001106.
\bibitem{adiabatic2} A. M. Childs, E. Farhi, and J. Preskill,  Phys. Rev. A {\bf 65}, 012322 (2001).
\bibitem{roland} J. Roland and N. J. Cerf, Phys. Rev. A \textbf{65}, 042308 (2002).
\bibitem{PRL} D. Daems and S. Gu\'erin, Phys. Rev. Lett. \textbf{99}, 170503 (2007).
\bibitem{STIRAP} N. V. Vitanov, T. Halfmann, B. W. Shore, and K. Bergmann,
Annu. Rev. Phys. Chem. \textbf{52}, 763 (2001).
\bibitem{Fleisch} M. Fleischhauer and M. D. Lukin, Phys. Rev. A \textbf{65}, 022314 (2002).
\bibitem{fstirap} N. V. Vitanov, K.-A. Suominen, and B. W. Shore, J. Phys. B {\bf 32}, 4535 (1999).
\bibitem{Berry} M. V. Berry, Proc. Roy. Soc. Lond. A \textbf{429}, 61 (1990).
\bibitem{Joye} A. Joye, C.-E. Pfister, J. Math. Phys. \textbf{34}, 454
(1993).
\bibitem{Elk} M. Elk, Phys. Rev. A \textbf{52}, 4017 (1995).
\bibitem{Drese} K. Drese and M. Holthaus, Eur. Phys. J. D \textbf{3}, 73 (1998).
\bibitem{Kuhn} S. Nussmann, M. Hijlkema, B. Weber, F. Rohde, G. Rempe, and A. Kuhn,
Phys. Rev. Lett. \textbf{95}, 173602 (2005).
\bibitem{fstirap_exp} V.A. Sautenkov, C.Y. Ye, Y.V. Rostovtsev, G.R. Welch, and M. O. Scully,
Phys. Rev. A \textbf{70}, 033406 (2004).
\bibitem{GJ} S. Gu\'erin and H. R. Jauslin, Adv. Chem. Phys. {\bf 125}, 147 (2003).
\end{thebibliography}
\end{document}